\documentclass[lettersize,10pt,comsoc]{IEEEtran}
\ifCLASSINFOpdf
\else
\fi
\usepackage{dsfont}
\usepackage{cite}
\usepackage{comment}
\usepackage{amsmath}
\usepackage{algorithm}
\usepackage{algorithmic}
\usepackage{graphicx}
\usepackage{color}
\usepackage{subfigure}
\usepackage{amssymb}

\begin{document}
%
\title{Interference Management for Full-Duplex ISAC in B5G/6G Networks: Architectures, Challenges, and Solutions}

\author{Aimin Tang,~\IEEEmembership{Member,~IEEE,} Xudong~Wang,~\IEEEmembership{Fellow,~IEEE,} and J.~Andrew~Zhang,~~\IEEEmembership{Senior Member,~IEEE}\\
\thanks{Aimin Tang and Xudong Wang are with the UM-SJTU Joint Institute, Shanghai Jiao Tong University, Shanghai, China. J.~Andrew~Zhang is with the School of Electrical and Data
Engineering, University of Technology Sydney, Australia.
Corresponding author: Xudong Wang, e-mail: wxudong@ieee.org.}
}

\maketitle

\begin{abstract}
Integrated sensing and communications (ISAC) has been visioned as a key technique for B5G/6G networks. To support monostatic sensing, a full-duplex radio is indispensable to extract echo signals from targets. Such a radio can also greatly improve network capacity via full-duplex communications. However, full-duplex radios in existing ISAC designs are mainly focused on wireless sensing, while the ability of full-duplex communications is usually ignored. In this article, we provide an overview of full-duplex ISAC (FD-ISAC), where a full-duplex radio is used for both wireless sensing and full-duplex communications in B5G/6G networks, with a focus on the fundamental interference management problem in such networks. First, different ISAC architectures are introduced, considering different full-duplex communication modes and wireless sensing modes. Next, the challenging issues of link-level interference and network-level interference are analyzed, illustrating a critical demand on interference management for FD-ISAC. Potential solutions to interference management are then reviewed from the perspective of radio architecture design, beamforming, mode selection, and resource allocation. The corresponding open problems are also highlighted.
\end{abstract}


%
\IEEEpeerreviewmaketitle

\section{Introduction}

Radio frequency (RF) can be used for both wireless communications and wireless sensing.
Besides the success in wireless communications, another prominent RF application is wireless sensing, which leverages wireless signals to sense targets in the surrounding environment of the transceivers. A typical wireless sensing system is radar system.
Since both systems utilize RF transceiver function blocks, fusing two systems into an integrated system, i.e., integrated sensing and communications (ISAC), has attracted great attention recently \cite{liu2022integrated}. ISAC holds several advantages. First, communications and sensing can reuse the spectrum so as to improve the spectrum utilization efficiency. Second, an ISAC system can reuse the hardware, reduce the cost, and minimize its size. Third, the communication and sensing functions can have a real-time interplay, and thus benefit each other to achieve a better performance. Thanks to the widely deployed communication infrastructures, integrating sensing with communication systems for perceptive mobile networks has been treated as one of the key technologies for Beyond 5G (B5G) and 6G networks \cite{zhang2020perceptive}.

\begin{figure*}
  \centering
  \includegraphics[width=0.9\linewidth]{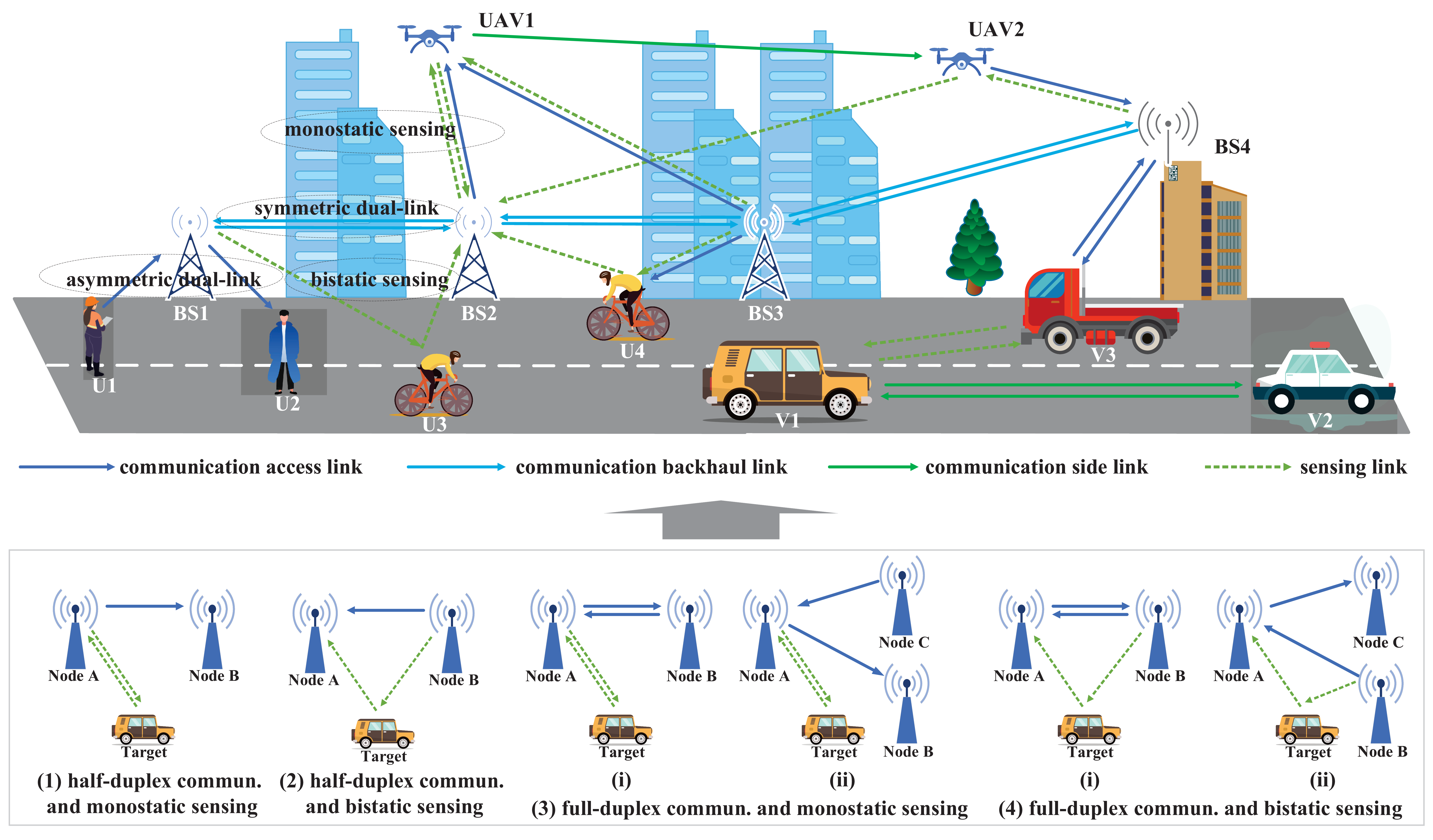}\\
  \vspace{-0em}
  \caption{Different ISAC architectures in B5G/6G networks. 
  }\label{fig:scenario}
  \vspace{-0em}
\end{figure*}

In order for a radio to transmit signals and receive echo signals for sensing at the same time, a full-duplex radio is the key enabler \cite{Barneto2021exp5G}. It can achieve simultaneous transmission and reception in the same frequency band \cite{bharadia2013full,singh2020millimeter}. Thus, it is also called in-band full-duplex radio.
For wireless communications, a full-duplex radio can either enable full-duplex communications to improve the link capacity for civilian communication networks, or be leveraged to conduct electronic warfare for enhancing military tactical communications \cite{riihonen2017inband}. Currently, full-duplex communications have already been investigated in the 3rd Generation Partnership Project (3GPP) Release 18 standard for B5G networks \cite{5gFD}. Therefore, integrating both wireless sensing and full-duplex communications with full-duplex radios is a natural choice for B5G/6G networks. In this article, full-duplex ISAC (FD-ISAC) is studied, where in-band full-duplex radios are considered for both communications and sensing.

A fundamental challenge in FD-ISAC is interference management.
It can be investigated at two levels: link-level and network-level. At the link-level, self-interference cancellation is the key challenge, and its requirements for sensing are different from those for communications. In full-duplex communications, all components of the transmitted signal (e.g., the direct leakage, and echoes from the environment) should be cancelled. However, in wireless sensing, direct leakage needs to be cancelled, while the echoes reflected from the targets are useful signals and should be retained as much as possible.
At the network-level, it is difficult to handle the cross-interference from the other transmission nodes. 
When sensing is not considered, full-duplex communications in a network level already confront the cross-interference management problem \cite{li2017full}. In FD-ISAC, since communications and sensing have different coverage ranges and sensitivities, the cross-interference management becomes more complicated. The link-level interference management for FD-ISAC has been investigated by some studies from some specific points such as radio architecture design in \cite{Barneto2021exp5G,tang2021self} and beamforming design in \cite{he2023full}, where each considers only one specific point. Moreover, network-level interference management for FD-ISAC is not investigated.

In this article, we provide a comprehensive overview on FD-ISAC systems, with a focus on interference management technologies. First, we introduce diverse FD-ISAC architectures, by combining different full-duplex communication links and sensing modes. These architectures exploit the facts that full-duplex radios can be utilized to support symmetric or asymmetric dual-link for full-duplex communications, and wireless sensing can be realized in monostatic and bistatic sensing modes. We then elaborate the challenges of interference management at both the link-level and the network-level.
We further review potential interference management technologies, from the perspective of radio architecture design, beamforming, mode selection, and resource allocation. Exemplified preliminary results are also presented, together with highlighted associated open research problems. Finally, concluding remarks are provided.

\section{Integrated Sensing and Full-Duplex Communications in B5G/6G networks}
In B5G/6G networks, there are generally two types of nodes: base station (BS) and user equipment (UE) as shown in Fig. \ref{fig:scenario}. Here, a UE can be a vehicle UE or unmanned aerial vehicle (UAV) UE. Both BS and UE can be a full-duplex radio.
Although the capability of full-duplex radios has been validated by several studies such as \cite{bharadia2013full,singh2020millimeter}, the complexity for self-interference cancellation is still very high, and thus full-duplex radios have not been widely deployed in current networks.
Therefore, it is also reasonable to assume only part of the nodes in B5G/6G networks are full-duplex radios \cite{li2017full}.

There are two types of full-duplex communication modes: symmetric dual-link such as the links between BS1 and BS2 in Fig. 1 and asymmetric dual-link such as the link U1-BS1-U2 in Fig. 1. In symmetric dual-link, both nodes are full-duplex radios. In asymmetric dual-link, the central node must be a full-duplex radio, and two end-nodes can be half-duplex. Due to the presence of half-duplex nodes in the network or traffic flow requirements, both symmetric dual-link and asymmetric dual-link need to be considered.
There are also two basic modes for wireless sensing: monostatic and bistatic sensing. In monostatic sensing, a node, e.g, BS2 in Fig. 1, transmits an illuminating signal and the echo signal from the target is then detected by the node itself. 
In bistatic sensing, a node, e.g., BS1 in Fig. 1, transmits an illuminating signal, but the echo signal from the target is detected by another node, e.g., BS2 in Fig. 1.
From a network perspective, both sensing modes can be leveraged to explore the sensing opportunities.

Considering communication modes and sensing modes, there are several ISAC architectures as shown in Fig. \ref{fig:scenario}.
\begin{itemize}
  \item First, half-duplex communications and monostatic sensing are enabled, where node A transmits communication signal to node B and also utilizes this communication signal for monostatic sensing.
  \item Second, half-duplex communications and bistatic sensing are enabled, where node A leverages the illuminating signal from node B for bistatic sensing. The illuminating signal from node B may be different in two cases. In one case, the data signal is for node A, so that node A can decode the data for further sensing processing. In the other case, the data signal is not for node A, so that node A can only leverage the public reference signal for further sensing processing due to the privacy issue.
  \item Third, full-duplex communications and monostatic sensing are enabled, where node A transmits communication signal to node B and also utilizes this signal for monostatic sensing. Moreover, node A also receives communication signal from node B or another node C.
  \item Fourth, full-duplex communications and bistatic sensing are enabled, where node A leverages the illuminating signal from node B for bistatic sensing and also transmits communication signal to node B or another node C. The illuminating signal may also be different for the two cases as shown in the second architecture.
\end{itemize}
All ISAC architectures can exist in B5G/6G networks as shown in Fig. \ref{fig:scenario}, depending on the deployment of full-duplex radios, real-time traffic load, and interference situations.

To support wireless sensing, the node with sensing function, i.e., ISAC full-duplex node, needs to have additional capability for processing sensing signal, such as more hardware/software sources. It also confronts more critical challenges of interference cancellation than that for full-duplex communications only. Therefore, in a practical deployment, only part of the BSs may be ISAC nodes in the network. For example, only the BS2 among the four BSs in Fig. \ref{fig:scenario} is set as an ISAC node. Therefore, the whole area for communications in Fig. 1 is covered by four BSs, while for sensing, the whole area is covered by only the BS2. In other words, BS2 needs to cover a larger area for sensing than that for communications. Thus, the sensing coverage of an ISAC BS is larger than its communication coverage, which will also incur new challenge of interference management for FD-ISAC.

\begin{figure}
  \centering
  \includegraphics[width=0.99\linewidth]{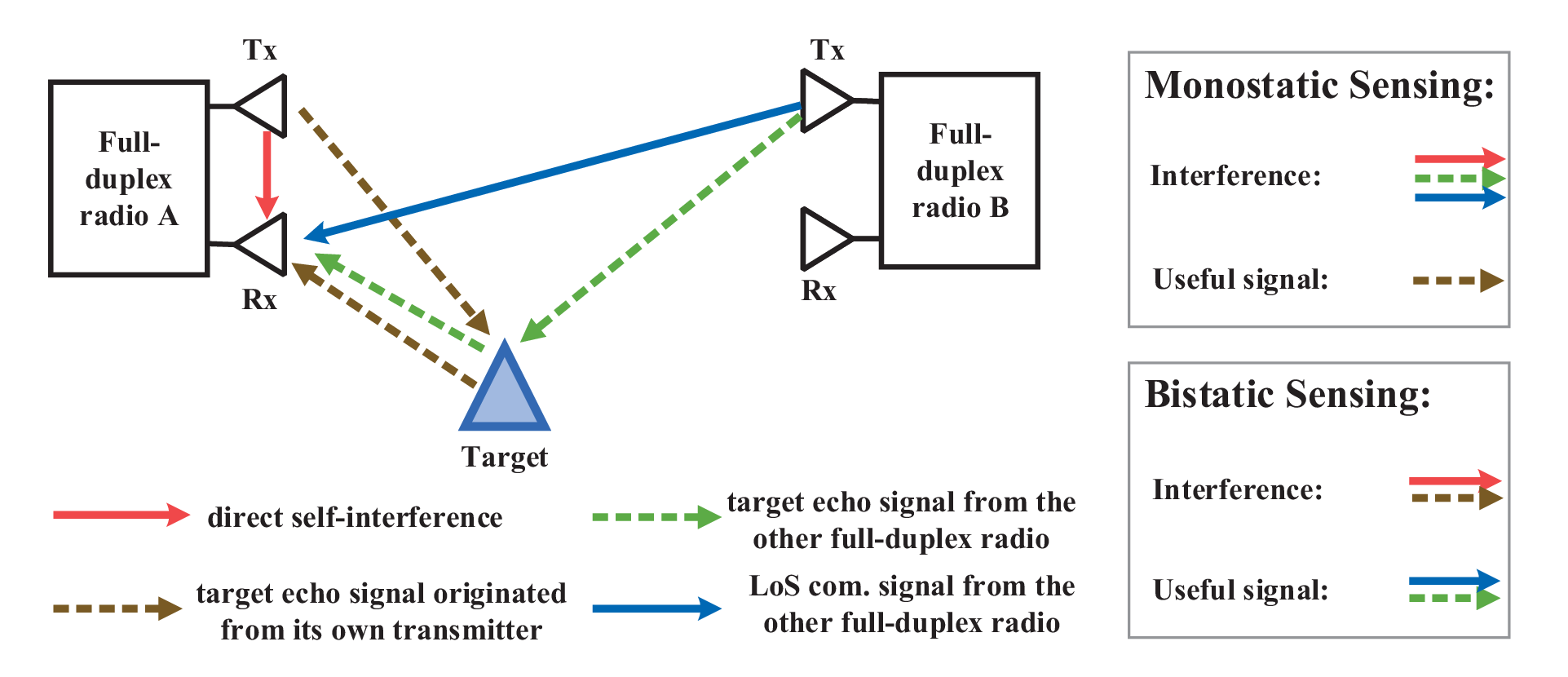}\\
  \caption{Illustration of link-level interference for FD-ISAC.}\label{fig:linkInterference}
  \vspace{-0em}
\end{figure}

\section{Challenges of Interference Management}

\subsection{Link-Level Interference for FD-ISAC}
For full-duplex communications, the self-interference consists of two main components: direct self-interference and echo self-interference, both of which need to be cancelled. The echo self-interference contains all the reflected signals from the environments.
Traditional full-duplex node for communications cancels both self-interference components \cite{bharadia2013full,singh2020millimeter}. 

In FD-ISAC, as shown in Fig. \ref{fig:linkInterference}, the received signal at the ISAC full-duplex radio (i.e., node A) consists of multiple components: the direct self-interference, the target echo signal originated from its own transmitter, and communication signals including echo signal and line-of-sight (LoS) path signal from the other full-duplex radio (i.e., node B).
The interference and useful signals for two sensing modes are separately analyzed as follows.

In the monostatic sensing mode, the direct self-interference needs to be cancelled for both full-duplex communications and sensing. The target echo signal originated from its own transmitter is an echo self-interference component for full-duplex communications, but is actually the useful signal for monostatic sensing. Thus, it should be cancelled for full-duplex communications, but retained for wireless sensing. The communication signal from the other full-duplex radio is also interference to wireless sensing. Since the communication signal usually has a signal-to-noise ratio (SNR) of 5-30 dB, the interference can significantly degrade the performance of sensing.

In the bistatic sensing mode, the target echo signal originated from node B is the desired signal for bistatic sensing. However, the LoS path signal is also useful, and can be either treated as a reference for system calibration \cite{li2023integrating} or used to decode the communication data for further sensing processing \cite{zheng2017super}. However, since the LoS path signal is usually much stronger than the target echo signal, it still needs to be cancelled for further target detection and estimation. Both the direct self-interference and echo signal originated from its own transmitter need to be cancelled for full-duplex communications and wireless sensing.

Based on the above analysis we can see that the interference cancellation requirements for full-duplex communications only and FD-ISAC are not fully consistent. Such an inconsistency requires new mechanisms to manage the interference for FD-ISAC.

\subsection{Network-Level Interference for FD-ISAC}

\begin{figure}
  \centering
  \includegraphics[width=0.9\linewidth]{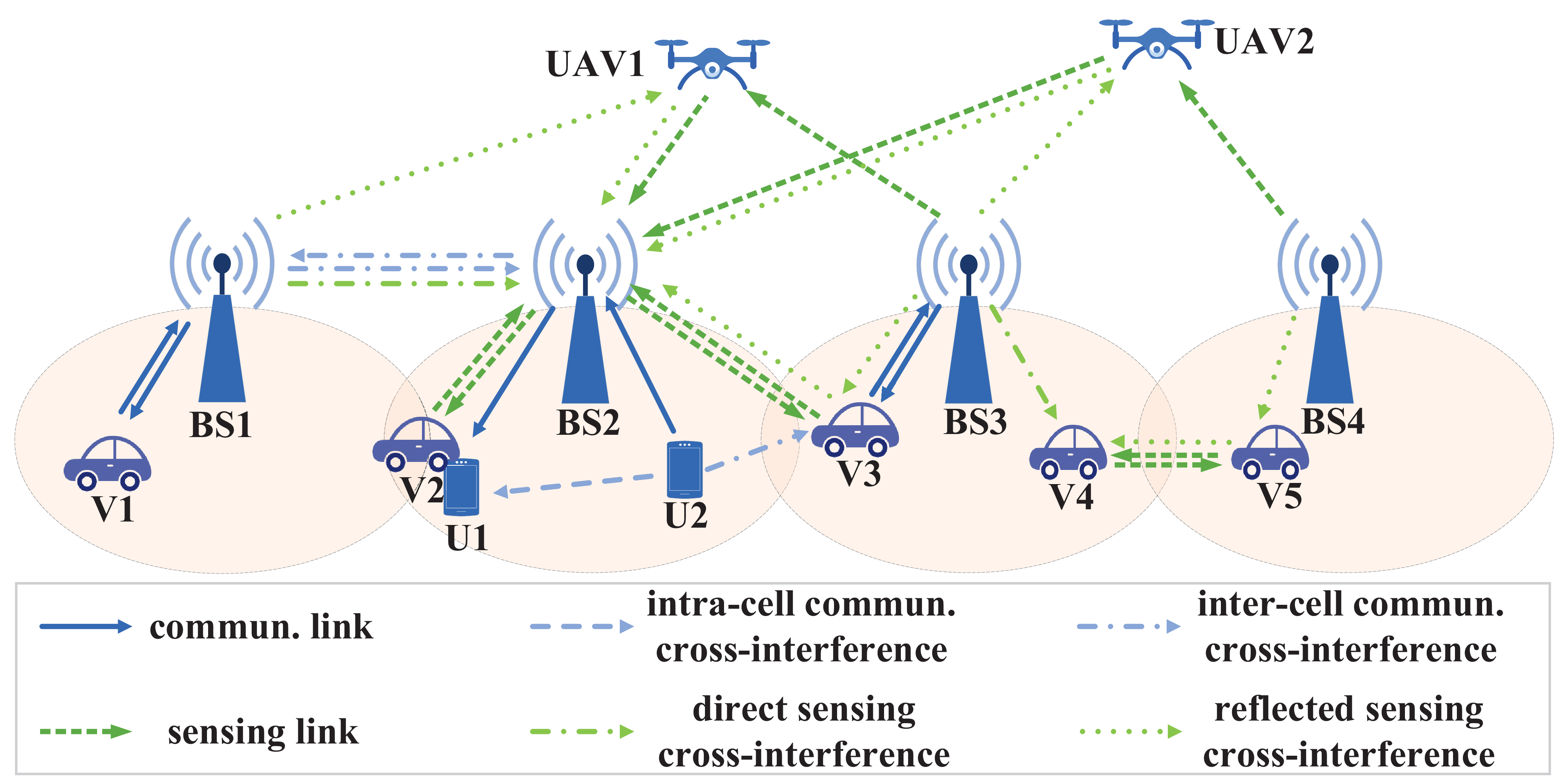}\\
  \caption{Illustration of network-level interference for FD-ISAC.}\label{fig:NetInterference}
  \vspace{-0em}
\end{figure}

For full-duplex communication networks, there are two types of network-level communication interference: intra-cell cross-interference and inter-cell cross-interference. When asymmetric dual-link is set, there may exist intra-cell communication cross-interference between two end-nodes such as the interference from U2 to U1 in Fig. \ref{fig:NetInterference}.
To boost the capacity of full-duplex networks, the cross-interference must be well managed or cancelled \cite{li2017full}.

In FD-ISAC, in addition to the network-level communication interference, the sensing interference can be classified in two types: direct cross-interference and reflected cross-interference. 
Compared to the communication function, the sensing function is more sensitive to interference strength. For example, the sensing detection can usually be achieved under an SNR even lower than -20 dB. Therefore, the ignored interference to communications may still have great impact on sensing. 
Refer to Fig. \ref{fig:NetInterference} for a better illustration of the sensing interference. When BS2 conducts monostatic sensing for V3, the interference signal from BS1 is the direct sensing cross-interference. 
The signal from the BS3-V3-BS2 link is the reflected sensing cross-interference from BS3. For communications, such a signal is usually ignored, since the reflected signal may be as weak as noise. 

If only part of the BSs are considered as the ISAC sensing nodes, the sensing coverage range will be much larger than the communication coverage. In such a case, the sensing interference will become more severe. For example, if BS2 leverages the illuminating signal from BS4 to achieve bistatic sensing for UAV2, the sensing interference from BS3 needs to be properly managed.

In short, from a network perspective, the cross-interference for both communications and sensing has to be properly managed. Since communications and sensing have different coverage ranges and sensitivities, additional interference management designs are required for FD-ISAC, compared to traditional full-duplex communication networks.

\section{Interference Management Technologies and Open Problems}

In this section, several potential technologies including radio architecture design, beamforming, mode selection, and resource allocation are proposed for interference management. Some of them, such as several modules designed in the radio architecture, are for link-level interference management, and some others, such as mode selection and resource allocation, are for network-level interference management. Beamforming can be used to address both link-level and network-level interference. 

\subsection{Radio Architecture Design}

\begin{figure}
  \centering
  \includegraphics[width=0.99\linewidth]{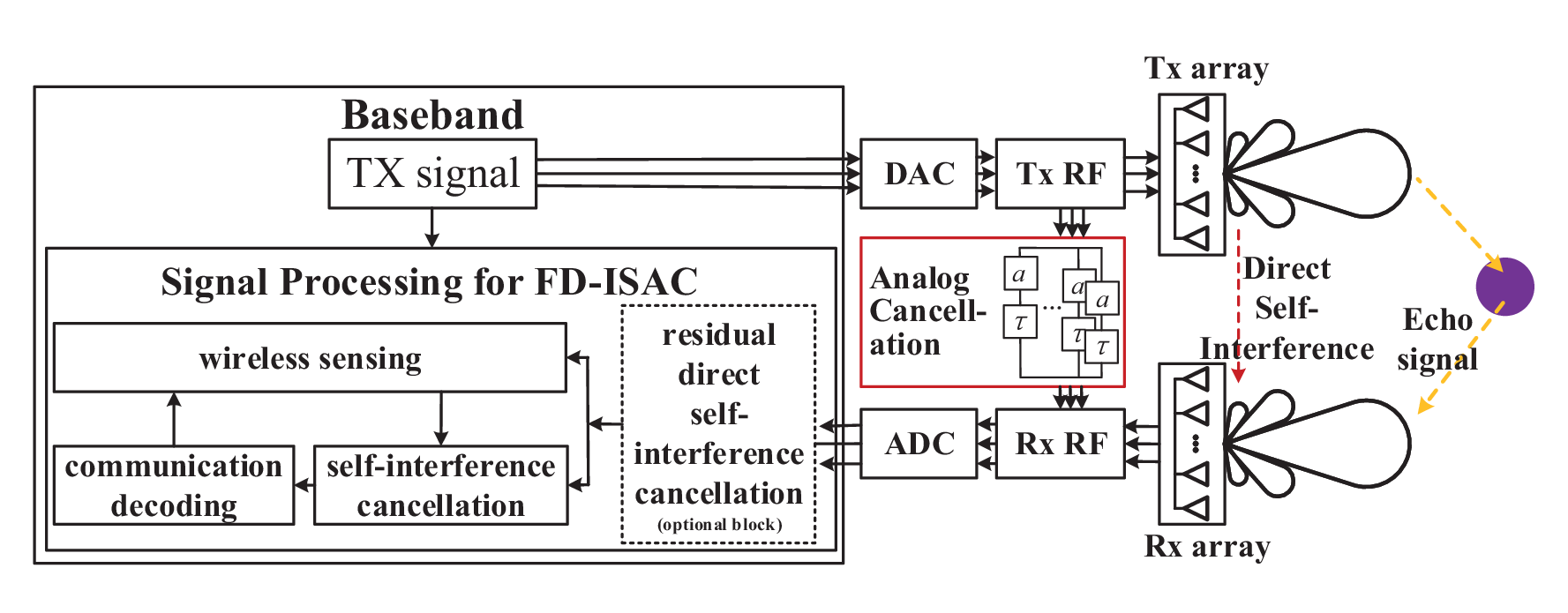}\\
  \vspace{-0em}
  \caption{The radio architecture design for an ISAC full-duplex node.}\label{fig:radioArch}
  \vspace{-0em}
\end{figure}

To handle the inconsistency of self-interference cancellation in FD-ISAC, a radio architecture for the ISAC full-duplex node is proposed as shown in Fig. \ref{fig:radioArch}.

The strong direct self-interference can saturate the analog-to-digital-converter (ADC). Thus, both antenna cancellation and analog cancellation for direct self-interference are required before ADC. Antenna cancellation can be through antenna separation and beamforming.
In the RF front-end, two separate antenna arrays can be used for transmit RF and receive RF. If there is sufficient space for deploying full-duplex radio at, e.g., BS, direct self-interference can be decreased by increasing the distance between transmit and receive antenna arrays. However, such a method may not be applicable at the user side. Beamforming is also an important way to achieve antenna cancellation as will be elaborated in the next subsection.
For analog cancellation circuits, several taps with fine-tuned attenuation and delay values are utilized. There are two techniques in the design of analog cancellation. First, to avoid to contaminate the echo signal, the fine-tuned delay is set to be lower than a threshold, since the delay of echo signal is usually much larger than that of the direct self-interference \cite{Barneto2021exp5G}. However, for ultra-short-range sensing, the design of analog cancellation taps is very challenging, since delays of direct self-interference and target echo are too close.
Second, as the antenna cancellation changes such as beampattern changes, the parameters in analog cancellation (and also digital cancellation) need to be adjusted accordingly. How to fast adapt the parameters to beampatterns is an open research problem, where machine learning may be useful to obtain the relationship.

In digital-signal-processing block, a digital self-interference cancellation module is firstly required to cancel the residual direct self-interference if it is not fully cancelled by the antenna suppression and analog cancellation. This digital cancellation module should be carefully designed so that only the direct self-interference is cancelled and the target echo signals are mostly retained. Such an algorithm design is a challenging open research problem. A few algorithms such as fixed digital calibration scheme \cite{tang2021self} and adaptive filter approach \cite{hassani2021joint} have been explored for such a digital cancellation.

To handle the different requirements of interference cancellation, a parallel signal processing can be designed for communications and sensing, as shown in Fig. \ref{fig:radioArch}. For wireless sensing, if monostatic sensing is considered, the transmitted digital signal fed from the transmission module is utilized as reference for sensing. For bistatic sensing, the decoded data from communication decoding module are used as reference for sensing.
As for communications, the sensing information of the target echo signals can be leveraged to assist self-interference cancellation for a better communication performance.

\subsection{Beamforming}
The beamforming technique can be used to handle both link-level and network-level interference, for either analog, digital, or hybrid beamforming. Since BSs usually have more antennas than end users, beamforming for interference management can be more efficiently implemented at the BS side. Two perspectives need to be considered in the beamforming design: beamforming gain and interference management.

High beamforming gain can be exploited to improve the signal-to-interference-plus-noise-ratio (SINR) for both communications and sensing.
Furthermore, beamforming also plays an important role for interference management in two aspects. First, when beamforming technique is applied, only the communication nodes or sensing targets in the mainlobe of the beam will get a high signal strength. Thus, the signal strength is naturally suppressed outside the mainlobe of the beam, which can limit the interference to a small angle coverage. Second, we can further design specific beampattern to suppress or null the signal in some specific angles, so as to limit the self-interference and cross-interference. For example, if separate antenna arrays are used for the ISAC full-duplex radio, the beamforming gain at the transmit (receive) antenna array can be designed to be nulled at the direction of receive (transmit) antenna array. Thus, the direct self-interference can be highly suppressed. 
At the same time, the beamforming gain at the desired direction for communications or sensing can be maximized via the optimization of beamformers \cite{he2023full}. The same method can be applied to cross-interference management.
Moreover, since digital signal processing for communication decoding and wireless sensing in the ISAC full-duplex receiver is conducted in parallel, separate receive digital beamformers can be applied to communications and sensing.

\begin{figure}
  \centering
  \includegraphics[width=0.99\linewidth]{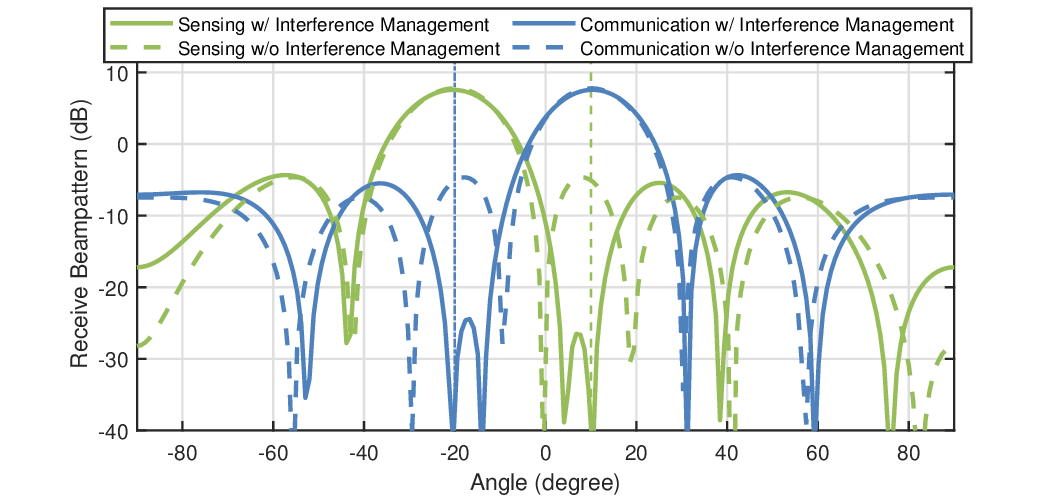}\\
  \vspace{-0em}
  \caption{Beampattern examples for receiver ISAC beamforming design.}\label{fig:beamforming}
  \vspace{-0em}
\end{figure}

An example of receive beamforming design for interference management with monostatic sensing is simulated. We assume the direct self-interference is already cancelled. Six RF chains are considered for receive beamforming design. We set the communication direction from node B to node A as $10^\circ$ and the target direction to node A as $-20^\circ$. The basic zero-forcing scheme, i.e., nulling the gain at the undesired direction, is used for interference management. The receive beampatterns for communications and sensing at node A are shown in Fig. \ref{fig:beamforming}. If interference management is not considered, the communication beampattern has a large sidelobe at the target direction, while the sensing beampattern has a large sidelobe at the communication direction. When interference management is applied, the receiver beampattern for communications can be nulled at the target direction to avoid the self-interference from target echo signal, while the receive beampattern for wireless sensing can be nulled at the communication direction to avoid cross-interference for sensing. Moreover, the beamforming gain for the desired direction is still maximized.

The above example shows the importance of interference management via beamforming from the receiver side with given directions. However, when the locations of communication receivers and targets are unknown or not accurate, how to conduct the beamforming design at both the transmitter and receiver is still a challenging problem. Furthermore, the beamforming design/optimization also depends on the SINR requirements of communications and sensing. Joint optimization of transmitter and receiver beamformers for both communications and sensing usually confronts high complexity. How to develop low-complexity algorithm for beamforming is also an open problem.

\subsection{Mode Selection}
In FD-ISAC networks, mode selection is crucial for interference management. Mode selection includes not only the selection of ISAC architecture, but also the pairing of user and target. For example, when the ISAC architecture with asymmetric full-duplex communications and monostatic sensing is considered, the user pairing of two end-users should be properly scheduled to avoid intra-cell cross-interference for communications. It is desired to pair a sensing target close to the communication user who receives the signal from the ISAC full-duplex node. In this case, the beamforming gain directions for communications and sensing are consistent, so that a better interference management result can be achieved. For example, when U1 and U2 are paired for asymmetric full-duplex communications in Fig. \ref{fig:NetInterference}, the target V2 that is close to U1 can be paired as the sensing target for monostatic sensing.

Furthermore, the selection between monostatic sensing and bistatic sensing is also highly related to the target range. Let us take the BS2 in Fig. \ref{fig:NetInterference} as the ISAC full-duplex node as an example. If the target is within the communication coverage of BS2, monostatic sensing may be a better choice than bistatic sensing. In this case, by pairing the sensing target with the communication users, it is much easier to control the interference within the communication coverage. However, if the sensing target is far away from BS2 such as two-hop away, bistatic sensing may be a better choice. In such a case, monostatic sensing usually needs to increase the transmission power for sensing the target due to a large two-way distance. The improved transmission power may lead to a large cross-interference for other communication links and sensing links.
For example, in Fig. \ref{fig:NetInterference}, the bistatic sensing link of BS4-UAV2-BS2 is a better choice than the monostatic sensing link of BS2-UAV2-BS2.

Achieving the optimal mode selections for interference management requires a global and centralized scheduling of the communication links and sensing links in the network, which is in general hard to implement in a large network. The communication link scheduling is also related to the real-time data traffic. Therefore, how to design an efficient mode selection scheme for FD-ISAC remains as an open problem.

\begin{figure}
  \centering
  \includegraphics[width=0.9\linewidth]{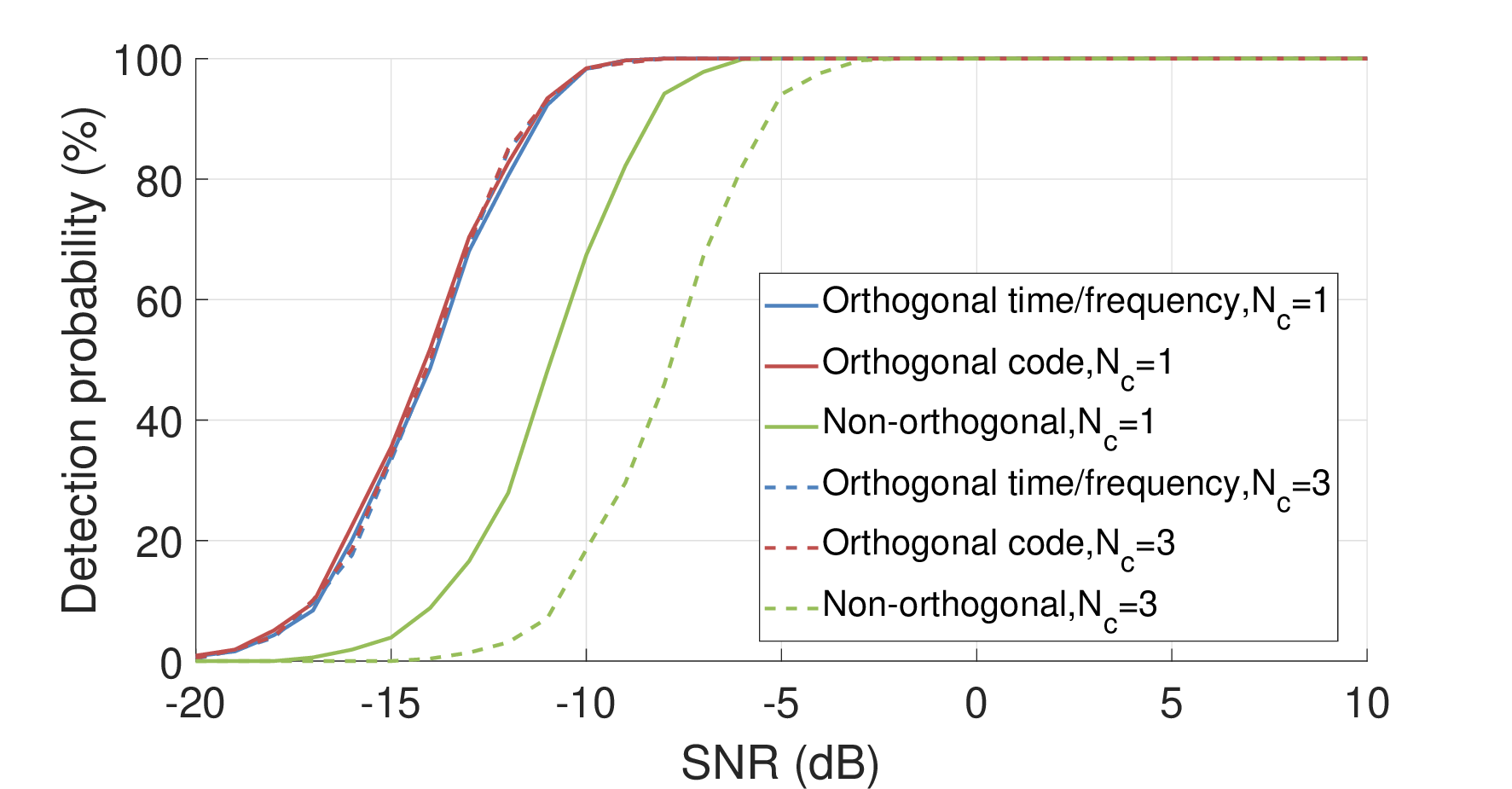}\\
  \vspace{-0em}
  \caption{Example of P/RS resource allocation for sensing detection probability.}\label{fig:signal}
  \vspace{-0em}
\end{figure}

\subsection{Resource Allocation}
Resource allocation is generally an efficient way to handle cross-interference, which has already been considered for full-duplex communication networks \cite{song2015resource}. In FD-ISAC, resource allocation in terms of time, frequency, spatial, and power resource can be jointly optimized to suppress the cross-interference for both communications and sensing, given the requirements of communications and sensing.

However, if only pilot/reference signal (P/RS) is utilized for sensing, resource allocation for P/RS can adopt some simple but effective strategies for managing cross-interference. First, the P/RS resources for different BSs can be designed with the same resource block, i.e., the sensing resource block is not overlapped with data resource block for different nodes. Second, since P/RS only occupies a small amount of resources, the P/RS can be allocated via orthogonal time/frequency or code to avoid or suppress the cross-interference. The P/RS density and power between P/RS and data can be further optimized by considering communication and sensing constraints \cite{zhao2023reference}.
An example showing the effect of resource allocation for P/RS is illustrated in Fig. \ref{fig:signal}, where a full-duplex BS is to sense a target, and there exists cross-interference from neighboring BSs.
We assume that the cross-interference from all neighboring BSs is the same, with a signal strength characterized by SNR of $\gamma=0$ dB. We simulate $N_\text{c}=1$ and $N_\text{c}=3$ neighboring BSs. The setup of system parameters follows 5G standard at 28 GHz with 100 MHz bandwidth. The constant false-alarm detector for radar sensing is used with false alarm rate set to $10^{-5}$.
The sensing detection probability with respect to target signal SNR is shown in Fig. \ref{fig:signal}. The orthogonal time/frequency P/RS design can avoid the cross-interference, and the orthogonal code design can achieve almost the same performance as orthogonal time/frequency. The performance gain of both designs over non-orthogonal P/RS resource allocation where the P/RS block for the sensing BS is allocated to be overlapped with the data block of the other BSs increases as the number of neighboring BSs increases, i.e., cross-interference increases.

Since resource allocation for FD-ISAC is coupled with many constraints such as beamforming and mode selection, different deployments of FD-ISAC require specific optimizations, tailored to the deployments.

\section{Conclusion}
This article provides an overview for FD-ISAC in B5G/6G networks, where full-duplex communications and wireless sensing are both enabled to fully explore the capability of full-duplex radios. Interference management in such a network becomes more challenging compared to that in conventional full-duplex communication networks. We provided a detailed review for the challenges and potential technologies of interference management. 

It is evident that the complicated interference in FD-ISAC cannot be effectively resolved by a single solution. Instead, it necessitates the integration of multiple solutions. However, these solutions can often be interdependent. For instance, modifying the beamforming pattern usually triggers the need of parameter update in self-interference cancellation. Consequently, it is insufficient to develop a solution to each open problem separately; instead, we need to consider such solutions in a systematical way so that FD-ISAC can be achieved practically. Furthermore, other techniques such as network architecture design and protocol design can also be incorporated to enhance interference management.

\section*{Acknowledgments}
This research work is supported by MediaTek Inc. The authors would like to thank the sponsor for the generous support.

\bibliographystyle{IEEEtran}
\bibliography{reference}

\begin{thebibliography}{10}
\providecommand{\url}[1]{#1}
\csname url@samestyle\endcsname
\providecommand{\newblock}{\relax}
\providecommand{\bibinfo}[2]{#2}
\providecommand{\BIBentrySTDinterwordspacing}{\spaceskip=0pt\relax}
\providecommand{\BIBentryALTinterwordstretchfactor}{4}
\providecommand{\BIBentryALTinterwordspacing}{\spaceskip=\fontdimen2\font plus
\BIBentryALTinterwordstretchfactor\fontdimen3\font minus
  \fontdimen4\font\relax}
\providecommand{\BIBforeignlanguage}[2]{{%
\expandafter\ifx\csname l@#1\endcsname\relax
\typeout{** WARNING: IEEEtran.bst: No hyphenation pattern has been}%
\typeout{** loaded for the language `#1'. Using the pattern for}%
\typeout{** the default language instead.}%
\else
\language=\csname l@#1\endcsname
\fi
#2}}
\providecommand{\BIBdecl}{\relax}
\BIBdecl

\bibitem{liu2022integrated}
F.~Liu, Y.~Cui, C.~Masouros, J.~Xu, T.~X. Han, Y.~C. Eldar, and S.~Buzzi,
  ``Integrated sensing and communications: Towards dual-functional wireless
  networks for {6G} and beyond,'' \emph{IEEE J. Sel. Areas Commun.}, vol.~40,
  no.~6, pp. 1728--1767, 2022.

\bibitem{zhang2020perceptive}
J.~A. Zhang, M.~L. Rahman, X.~Huang, Y.~J. Guo, S.~Chen, and R.~W. Heath,
  ``Perceptive mobile networks: Cellular networks with radio vision via joint
  communication and radar sensing,'' \emph{IEEE Veh. Tech. Mag.}, vol.~16,
  no.~2, pp. 20--30, 2020.

\bibitem{Barneto2021exp5G}
C.~B. Barneto, S.~D. Liyanaarachchi, M.~Heino, T.~Riihonen, and M.~Valkama,
  ``Full duplex radio/radar technology: The enabler for advanced joint
  communication and sensing,'' \emph{IEEE Wireless Commun.}, vol.~28, no.~1,
  pp. 82--88, 2021.

\bibitem{bharadia2013full}
D.~Bharadia, E.~McMilin, and S.~Katti, ``Full duplex radios,'' in \emph{Proc.
  ACM SIGCOMM}, 2013, pp. 375--386.

\bibitem{singh2020millimeter}
V.~Singh, S.~Mondal, A.~Gadre, M.~Srivastava, J.~Paramesh, and S.~Kumar,
  ``Millimeter-wave full duplex radios,'' in \emph{Proc. ACM Mobicom}, 2020,
  pp. 1--14.

\bibitem{riihonen2017inband}
T.~Riihonen, D.~Korpi, O.~Rantula, H.~Rantanen, T.~Saarelainen, and M.~Valkama,
  ``Inband full-duplex radio transceivers: A paradigm shift in tactical
  communications and electronic warfare?'' \emph{IEEE Commun. Mag.}, vol.~55,
  no.~10, pp. 30--36, 2017.

\bibitem{5gFD}
3rd Generation Partnership Project~(3GPP), ``{Study on Evolution of NR Duplex
  Operation ({Release 18})},'' \emph{Technical Report (TR) 38.858}, 2022.

\bibitem{li2017full}
R.~Li, Y.~Chen, G.~Y. Li, and G.~Liu, ``Full-duplex cellular networks,''
  \emph{IEEE Commun. Mag.}, vol.~55, no.~4, pp. 184--191, 2017.

\bibitem{tang2021self}
A.~Tang, S.~Li, and X.~Wang, ``Self-interference-resistant {IEEE} 802.11
  ad-based joint communication and automotive radar design,'' \emph{IEEE J.
  Sel. Top. Signal Process.}, vol.~15, no.~6, pp. 1484--1499, 2021.

\bibitem{he2023full}
Z.~He, W.~Xu, H.~Shen, D.~W.~K. Ng, Y.~C. Eldar, and X.~You, ``Full-duplex
  communication for {ISAC}: Joint beamforming and power optimization,''
  \emph{IEEE J. Sel. Areas Commun.}, 2023.

\bibitem{li2023integrating}
S.~Li, C.~Luo, A.~Tang, X.~Wang, C.~Xu, F.~Gao, and L.~Cai, ``Integrating
  passive bistatic sensing into mmwave {B5G/6G} networks: Design and experiment
  measurement,'' in \emph{Proc. IEEE ICC}, 2023, pp. 2952--2957.

\bibitem{zheng2017super}
L.~Zheng and X.~Wang, ``Super-resolution delay-doppler estimation for {OFDM}
  passive radar,'' \emph{IEEE Transa. Signal Process.}, vol.~65, no.~9, pp.
  2197--2210, 2017.

\bibitem{hassani2021joint}
S.~A. Hassani, B.~van Liempd, A.~Bourdoux, F.~Horlin, and S.~Pollin, ``Joint
  in-band full-duplex communication and radar processing,'' \emph{IEEE Sys.
  J.}, vol.~16, no.~2, pp. 3391--3399, 2021.

\bibitem{song2015resource}
L.~Song, Y.~Li, and Z.~Han, ``Resource allocation in full-duplex communications
  for future wireless networks,'' \emph{IEEE Wireless Commun.}, vol.~22, no.~4,
  pp. 88--96, 2015.

\bibitem{zhao2023reference}
Q.~Zhao, A.~Tang, and X.~Wang, ``Reference signal design and power optimization
  for energy-efficient {5G V2X} integrated sensing and communications,''
  \emph{IEEE Trans. Green Comm. Netw.}, vol.~7, no.~1, pp. 379--392, 2023.

\end{thebibliography}

\begin{IEEEbiography}[{\includegraphics[width=1in,height=1.25in,clip,keepaspectratio]{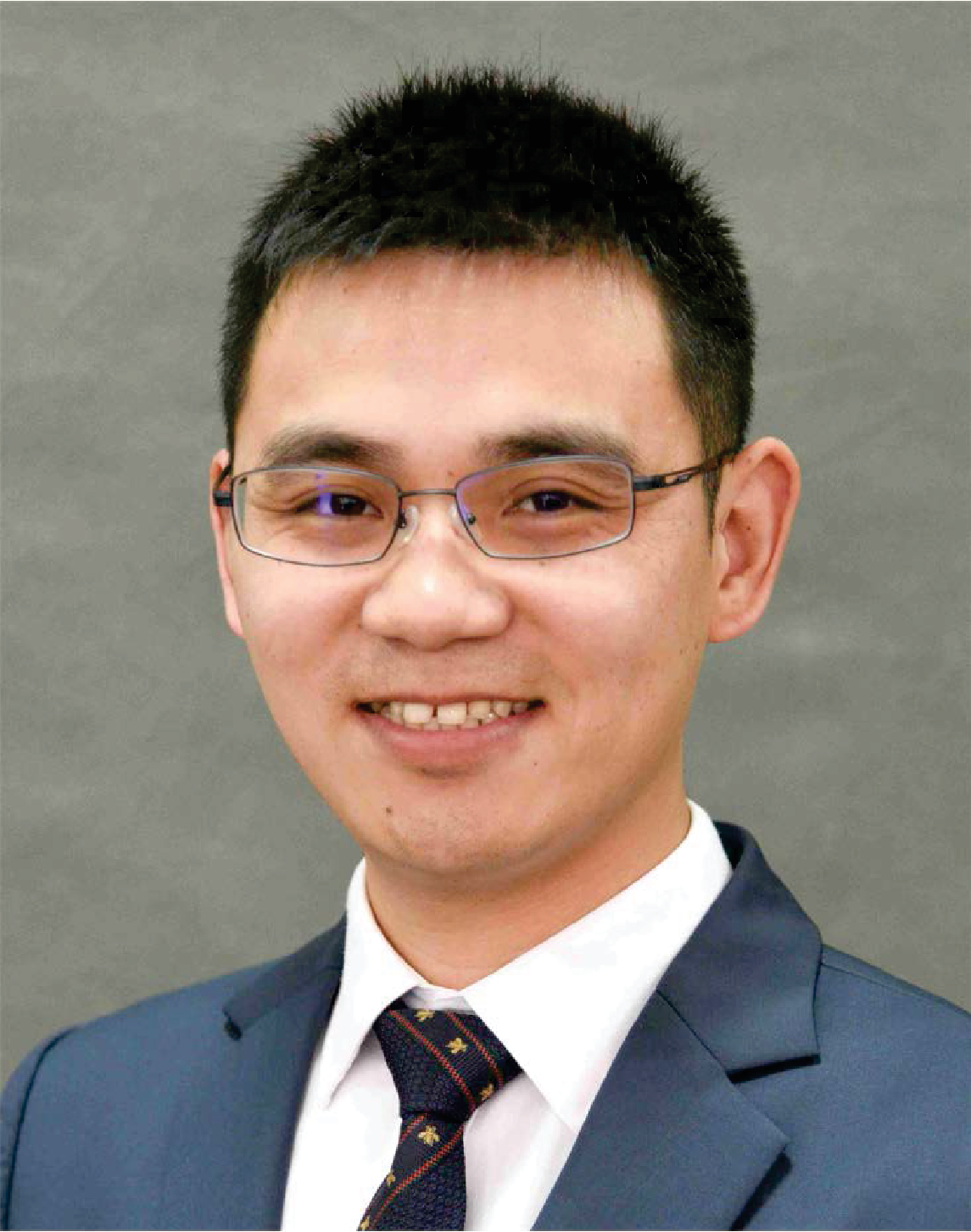}}]{Aimin Tang}
(Member, IEEE) received the B.S. and Ph.D. degrees in information and communication engineering from Shanghai Jiao Tong University, Shanghai, China, in 2013 and 2018, respectively.
He is currently a Research Assistant Professor with the University of Michigan-Shanghai Jiao Tong University (UM-SJTU) Joint Institute, Shanghai Jiao Tong University. His current research interests include B5G/6G networks, integrated sensing and communications, and full-duplex communications. 
\end{IEEEbiography}

\begin{IEEEbiography}[{\includegraphics[width=1in,height=1.25in,keepaspectratio]{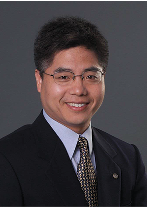}}]
{Xudong Wang} (Fellow, IEEE) received the Ph.D. degree in electrical and computer engineering from the Georgia Institute of Technology in 2003. He is currently the John Wu and Jane Sun Chair Professor of engineering at the UM-SJTU Joint Institute, Shanghai Jiao Tong University. He is also an affiliate Professor with the Department of Electrical and Computer Engineering, University of Washington. His research interests include wireless communication networks, distributed machine learning, and joint communications and sensing. 
\end{IEEEbiography}

\begin{IEEEbiography}[{\includegraphics[width=1in,height=1.25in,keepaspectratio]{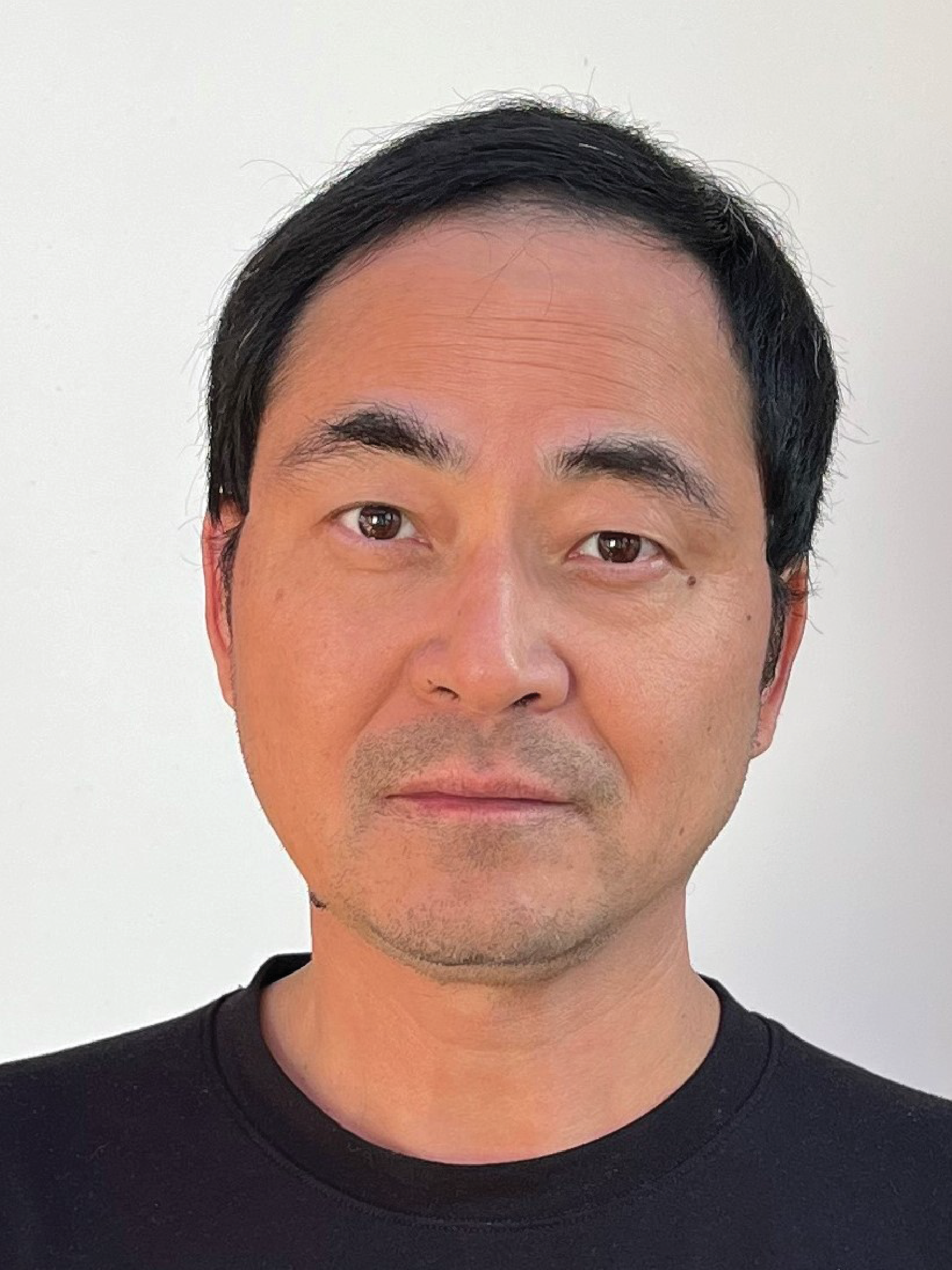}}]
{J. Andrew Zhang} (Senior Member, IEEE) received his M.Sc. degree from Nanjing University of Posts and Telecommunications, China, and his Ph.D. degree from the Australian National University, Canberra, in 1999 and 2004, respectively. Currently, he is a Professor in the School of Electrical and Data Engineering, the University of Technology Sydney, Australia. His research interests include signal processing for wireless communications and sensing and autonomous vehicular networks. 
\end{IEEEbiography}

\end{document}